# How Platform-User Power Relations Shape Algorithmic Accountability: A Case Study of Instant Loan Platforms and Financially Stressed Users in India

Divya Ramesh*
University of Michigan
Ann Arbor, USA
dramesh@umich.edu

Vaishnav Kameswaran
University of Michigan
Ann Arbor, USA
vaikam@umich.edu

Ding Wang
Google Research
Bangalore, India
drdw@google.com

Nithya Sambasivan
Unaffiliated
San Francisco, USA
nithyas@gmail.com

## ABSTRACT

Accountability, a requisite for responsible AI, can be facilitated through transparency mechanisms such as audits and explainability. However, prior work suggests that the success of these mechanisms may be limited to Global North contexts; understanding the limitations of current interventions in varied socio-political conditions is crucial to help policymakers facilitate wider accountability. To do so, we examined the mediation of accountability in the existing interactions between vulnerable users and a 'high-risk' AI system in a Global South setting. We report on a qualitative study with 29 financially-stressed users of instant loan platforms in India. We found that users experienced intense feelings of indebtedness for the 'boon' of instant loans, and perceived huge obligations towards loan platforms. Users fulfilled obligations by accepting harsh terms and conditions, over-sharing sensitive data, and paying high fees to unknown and unverified lenders. Users demonstrated a dependence on loan platforms by persisting with such behaviors despite risks of harms such as abuse, recurring debts, discrimination, privacy harms, and self-harm to them. Instead of being enraged with loan platforms, users assumed responsibility for their negative experiences, thus releasing the high-powered loan platforms from accountability obligations. We argue that accountability is shaped by platform-user power relations, and urge caution to policymakers in adopting a purely technical approach to fostering algorithmic accountability. Instead, we call for situated interventions that enhance agency of users, enable meaningful transparency, reconfigure designer-user relations, and prompt a critical reflection in practitioners towards wider accountability. We conclude with implications for responsibly deploying AI in FinTech applications in India and beyond.

## 1 INTRODUCTION

Accountability is necessary to ensure that artificial intelligence (AI) is deployed responsibly, especially given the wide applicability of AI algorithms to several automated decision making contexts with 'high stakes' [40, 57, 58, 81]. While automated decision systems (ADS) [106] have the potential to make more efficient and fairer decisions than their human counterparts [45, 52], they could also produce harmful outcomes, worsening inequality in society [20, 23, 50, 54, 92, 94, 98]. Through accountability relationships, the actors responsible for harms caused by the ADS can be obligated to provide 'accounts' to the individuals who are harmed; the individuals or their representatives may then judge the accounts, and seek to impose consequences if necessary [121]. In this way, we can ensure that the use of ADS occurs in accordance with the interests of all stakeholders. Facilitating organizational and technical transparency could reduce distrust among stakeholders, and enhance accountability relationships [10, 53, 90, 112]. Given their success in the US and UK, transparency mechanisms such as audits and explainability are being mandated in policies worldwide [9, 28, 91]. Consequently, information disclosure by technology providers[1] is often viewed as a precursor for algorithmic accountability [53, 64, 89].

---
[1] governments and corporations overseeing the design, development, deployment and/or procurement of ADS

However, recent work suggests that the success of enhancing accountability relationships through transparency may be limited. Perceived agency of stakeholders [69, 81], their education levels [43], and their optimism in AI [73], could complicate the rhetoric of 'stakeholder distrust in ADS.' Further, the efficacy of transparency mechanisms towards accountability depends on the presence of a critically-aware public, legislative support, watchdog journalism, and the responsiveness of technology providers [17, 60, 76]. Unfortunately, these preconditions may be unique to Global North contexts [109]. Understanding the limitations of current approaches in varied socio-political conditions is crucial to help policymakers adopt context-appropriate interventions, and ensure wider accountability. Prior work has sought to ease the burden on technology providers towards fulfilling transparency obligations [89, 102], and studied their impacts on users affected by ADS [42, 113, 122]. Yet, we know little about on-the-ground manifestations of accountability in ecosystems where some of its preconditions do not hold true.

To fill this gap, we examined how algorithmic accountability is mediated in existing interactions between vulnerable users and a 'high-risk' ADS in a Global South setting; one where there is weak legislation and nation-wide high optimism for AI. We conducted a qualitative study with financially stressed low and middle income users of instant loan platforms in India. These platforms target 'thin-file' borrowers (i.e., users ineligible to offerings from formal financial services) with various small credit offerings, often in the range of INR 500 - INR 100,000 (USD 7 - USD 1500). The loan platforms use machine learning algorithms trained on alternative data[2] to model risk and make lending decisions [5]. Instant loan platforms have risen to prominence in recent years through a combination of factors such as affordable smartphones [80], the state's push for widespread digital adoption [97], promotion of financial technology (FinTech) as the poster child of AI success in India [11], and financial challenges to users brought by the COVID-19 pandemic [18].

Through semi-structured interviews with 29 users of instant loan platforms from low and middle income groups in India, we examined how financially stressed users made meaning of their experiences with the 'high-risk' ADS, and how they perceived their relations to accountability. We found that users were drawn to loan platforms due to the promises of immediate money, minimal verification, and long tenure periods, which were enabled by instantaneous and synchronous aspects of AI. Users also perceived additional benefits such as enhanced privacy and dignity, preserved social ties, and social mobility through the use of these platforms. Since users had few avenues to seek financial assistance, they perceived instant loans as 'boons', and developed emotional attachments towards lenders. Users perceived and fulfilled several obligations towards lenders, even at the risks of undergoing abuse, discrimination, emotional and reputation harms, and self-harm from them. Yet, instead of being enraged with loan platforms, users shared responsibility for their negative experiences.

Through this work, we make the following contributions: First, we explore the relationship between ADS experiences of users, their social conditions and accountability. In doing so, we build upon previous work in FAccT, and explore the social dimensions of accountability through a case study on loan platforms in the Global South. We make an empirical contribution regarding how low-powered users perceive and demonstrate a dependence on the 'high-risk' ADS, holding themselves responsible for its failures; and how these user behaviors release high-powered actors from accountability obligations. Next, situating our findings in the literature on accountability, we argue that algorithmic accountability is mediated through platform-user power relations, and can be inhibited by socio-political realities of the context. We urge caution to policymakers in adopting universal technical interventions to foster accountability, and instead propose situated [114] approaches towards achieving parity in platform-user power relations. Our proposal includes: 1) Enhancing user agency through critical awareness, 2) Enabling meaningful transparency through collective spaces, 3) Re-configuring designer-user relations through community engagement, and 4) Committing to justice through critical reflection. We conclude with implications for using 'alternative data' in FinTech applications in the Global South.

## 2 RELATED WORK

In this section, we give a brief overview of the platform-user dynamics envisioned in literature on algorithmic accountability, the mechanisms designed to structure accountability relationships, and a glimpse into India's AI landscape.

### 2.1 Platform-User Dynamics in Algorithmic Accountability

Technical and organizational opacity are viewed as primary barriers to fostering accountability [27, 98], suggesting the need for transparency from technology providers [35–37]. Technology providers have also taken steps towards increasing transparency due to a combination of user demands and self-imposed responsibility [19, 84, 117, 123]. For instance, researchers studying the experiences of ADS among users have reported increased distrust among users and their desire for transparency into ADS [25, 47], Uber drivers accused the company of deception due to its use of an opaque ADS and demanded more transparency [116], and Yelp users expressed the need for transparency into its recommendation algorithm [48]. Consequently, several regulatory policies mandate public access to information, hoping that affected users will use this information towards making accountability demands from technology providers [16, 32, 51, 59, 99]. In fact, such an approach has been extremely successful recently. After audit trails of harmful facial recognition systems were made available to the public, there were widespread public campaigns that eventually led to their regulation in the US and UK [26]. Twitter took steps to modify its biased image cropping algorithm to satisfy user demands [123]. However, recent work casts a doubt on the generalizability of these platform-user dynamics to varied contexts. Favorable outcomes to users from otherwise discriminatory ADS may impede their accountability efforts [48, 120], users may resist imputing moral responsibility to ADS [21], and notions of accountability could vary by users' backgrounds [70]. In addition, platform-user relations may be more nuanced than the often cited rhetoric of distrust. Nation-states and users in the Global South view ADS aspirationally and deferentially [109]; where users may attribute far reaching capabilities to ADS, placing misguided trust in them [93]; and ADS may enjoy a legitimized power to influence users' actions, even with little or no evidence of their true capabilities [73]. Prior work calls for aligning transparency with user needs [42, 75]. These findings warrant a closer examination of the accountability dynamics in varied socio-political conditions; a gap that we seek to fill in this paper.

---
[2] non-traditional financial modeling data such as mobile phone and social media usage, financial transactions, images and videos used to model risk

## 2.2 Mechanisms of Algorithmic Accountability

A recent survey of algorithmic accountability policies in the public sector from 20 national and local governments found that transparency was the prime focus of policies [9]. Under a dynamic where users express skepticism, and seek to take action towards accountability, transparency (i.e., of models, datasets and practices surrounding the development of ADS) are viable mechanisms [53, 64, 89]. Mechanisms to increase transparency can be standalone such as *documentation* (i.e., of source-code, datasets, models, and processes surrounding the development of ADS) [21, 30, 55, 64, 83, 105] and *explainable decisions* (i.e., to help the users make informed choices when interacting with ADS) [10, 21, 42, 83, 100, 119]; or be embedded in other mechanisms such as *algorithmic audits* [26, 46, 112] and *impact assessments* [101, 102]. In fact, some studies with audits and explainability mechanisms have documented positive outcomes such as raising users' critical awareness [100], increasing their desires to seek accountability from the designers of ADS [113], and influencing technology providers to make changes in ADS [46, 101]. However, the efficacy of mechanisms in fostering accountability also relies on other factors such as a critically-aware public, legislative support, watchdog journalism, and the responsiveness of high-powered actors [17, 60, 76]. Raji and Buolamwini acknowledge the importance of consumer awareness and capitalistic competition in complementing their audit efforts in facial recognition regulation [101]. Unfortunately, such surrounding conditions for accountability are not universally available [109]. Organizational-level changes from technology providers most often occur as a result of regulatory and user pressures [103], low-powered users may find it challenging to regain their agency displaced by platforms [69, 81], and mechanisms may have limited efficacy where there is power asymmetry [17, 76]. This line of work calls for examining platform-user power relations for designing mechanisms.

## 2.3 AI Landscape in India

India, a country with 1.38 billion where half the population is under the age of 25, is considered an emerging force in AI due to its growing information technology workforce, research in AI, investments and cloud-computing infrastructure [31]. India envisions AI as a force for socio-economic upliftment, which is seen through state-supported industry initiatives [49] and wide deployment of AI in surveillance [67], agriculture [34], and welfare processing systems [49]. However, such promotion is supported with weak legislation. The two national AI strategies i.e., the AI Task force report [115], and the NITI Aayog's National Strategy for AI [91] are focused on increasing adoption of AI [115] or include prescriptive guidelines towards accountability with insufficient enforcement mechanisms [91]. Similar recommendations are found in state-level policies on AI in India [56]. Policy-oriented research from the FAccT and HCI communities have pointed out how adopting accountability frameworks from the Global North may fail without due consideration to the local contexts where they are applied [71, 85, 109]; Sambasivan et al., have noted the differences in axes discrimination and notions of fairness [109]. Kalyanakrishnan et al. and Marda et al., documented the amplification of biases when using Western frameworks in the Indian context [71, 86]. We contribute to this emerging line of work on AI policy and research agenda in India.

## 3 METHODS

### 3.1 Interviews

We conducted 29 semi-structured interviews with low and middle income individuals (16 men, 13 women) primarily from Karnataka and Tamil Nadu regions in India. We recruited participants who had used instant loan apps from non-banking financial companies between 6 months and 2 years of our study through DoWell Research agency and snowball sampling. We provided INR 1500 (USD 20) as incentives to our participants. We sampled participants based on age, gender, prior experience using instant loan applications and success of loan approval. We conducted virtual interviews in English, Kannada, Tamil and Hindi lasting 35-110 minutes (average of 55 minutes). The first author conducted 26 interviews (2 with the help of a translator), and a non-author colleague conducted 3 interviews. We sought prior written consent, and informed verbal consent before the start of the interviews. During the interviews, we focused on eliciting narratives [79] from participants to understand 1) their interests, their education and family backgrounds, 2) their financial situations during the pandemic, 3) their experiences with lending and borrowing through instant loan apps and other means, and 4) their notions of justice in lending and borrowing. Interviews were transcribed and/or translated within 2-3 days of each interviews. We used a professional service for transcribing the regional language interviews, which were all then individually verified by the first author. Towards the end of the interviews, we used a scenario as a probe to elicit participants' opinions on alternative credit. We first explained what AI meant to participants' through examples of Youtube and Facebook, and then presented this scenario: *Due to COVID, many people are in need of money but don't have jobs, or access to PAN cards and bank accounts. Some apps suggest using AI to make lending decisions. Instead of bank details, they will look at users' mobile phone information such as biometrics, location, call logs, financial transactions and shopping apps used on devices, and users' social media activity to make decisions on loan applications. They believe that this approach will increase people's access to loans. What are your thoughts about this?*

### 3.2 Analysis

We conducted reflexive thematic analysis to analyze our data [24]. In the familiarization phase, the first author listened to each audio recording at least once, and read each transcript at least twice, paying close attention to participants' choice of phrases, especially in regional languages, their emotional reactions to questions, hesitations, pauses, and repetitions. We recorded these observations and reflections and shared them during weekly research meetings with the rest of the team, which then served as aids in coding the data. In the coding phase, the lead author followed an open-coding approach first, staying close to the data (i.e., *needing money urgently, not telling friends the reason for money*) [108], and iteratively revised the codes with the second author (i.e., *'instant' money, preserving privacy, feelings of indebtedness*), resolving disagreements through discussion. We generated and refined themes by going over the data, engaging with literature, and through weekly research meetings with third and fourth authors. In this work, we present 3 themes that we generated from 11 stable codes: (1) Perceived Benefits of Instant Loan Platforms, (2) Perceived Obligations to Instant Loan Platforms, and (3) Dependence on Instant Loan Platforms.

### 3.3 Ethical Considerations and Limitations

We approached this topic with great care, knowing the dire circumstances of participants. We reflected carefully if this study was time-appropriate. Several participants were ecstatic to be part of our research to express their gratitude through our report towards loan companies. One participant requested extra time to share their experiences in depth. These incidents helped us viewed our participants as individuals in their own right, rather than as victims of their circumstances, and gave us confidence that this research was timely. During the interview, we let participants guide the discussion towards the experiences that were most salient to them. We stored data on Google drive and restricted access to the research team. We also took care to anonymize the data and report them in this paper. We intentionally do not specify the names of loan platforms that we recruited users from to preserve anonymity. Although we attempted to recruit participants across gender, our sample skews more towards men. We also do not have any perspectives from non-binary identifying individuals. Due to the COVID-19 pandemic, we conducted all our interviews over video and phone, which limited our ability to include observations and contextual inquiry. The first author's caste and class privilege (evident through name and dialect) may have influenced participants' responses. Our research team has over 10 years of experience working with marginalized populations in the Global South. Reflecting on our positionality, we elicited narratives with care, and analyzed the data extensively to cover multiple themes, and ensure validity.

## 4 CONTEXT
### 4.1 Participant Demographics

10 participants belonged to urban-middle income groups, and the rest 19 participants belonged to urban-low or lower-middle income groups. 25 of our participants worked in the service sectors as accountants and chefs in restaurants, carpenters, customer service, sales and marketing representatives, tailors, taxi and auto-rickshaw drivers, or owned small businesses. 2 participants worked in health and education sectors, and 2 participants identified their primary roles as "house-wives." All our participants incurred significant loss of incomes during the pandemic. Several participants (n=16) were responsible for supporting 4-5 member households with reduced or no incomes; they had pledged or sold the few assets they possessed, and in few cases, the very assets that were sources of income to them. In addition, vulnerability for them meant having to comply with exploitative rules from informal lenders, from their children's schools, from local state offices, and participants having no monetary or social capital to even claim their rights.

### 4.2 Instant Loan Applications

Instant loan platforms, primarily classified as 'FinTech' provide technical infrastructure to connect NBFCs (shadow banking entities that offer financial services without a banking license) [1] with borrowers. They offer small, short term loans, typically INR 500 - 500,000 over a period of 15 days - 6 months, using machine learning on a combination of CIBIL scores[3], and 'alternative credit data.' Although the workings of these loan apps are proprietary, most instant loan apps, in their privacy policy, disclose using the following as 'alternative credit data': 'know-your-customer' (KYC) data such as names, addresses, phone numbers, PIN codes, reference contacts, photos, and videos, personal account number (PAN), Aadhar number (unique identification number); device information such as location, hardware model, build model, RAM, storage, unique device identifiers like Advertising ID, SIM information that includes network operator, WIFI and mobile network information, cookies; financial SMS sent by 6-digit alphanumeric senders, and information obtained from 3rd party providers for making credit decisions [2, 4–7]. Applications also use this data for analyzing user behavior for advertising and security purposes. Apps use AI in several other ways including use facial recognition for completing verification, natural language processing for information extraction and contract automation, machine learning for fraud detection and market analysis, and chatbots to provide customer service [3]. These platforms, targeted at borrowers from low and middle income groups, have proliferated the market recently, and are hailed by the state as the 'drivers of economic growth' for the 'unbanked' India [11]. Some popular apps include Kissht [5], Dhani [2], KreditBee [6], SmartCoin [7], and MoneyTap [4].

## 5 FINDINGS

All participants unanimously cited the promise of 'instant cash' as primary reasons for trying instant loan platforms. We found that this promise was the precursor to a cycle of reciprocal exchanges between loan platforms and the users, which we discuss with the help of the following themes: (1) Perceived Benefits of Instant Loan Platforms, (2) Perceived Obligations to Instant Loan Platform, and (3) Dependence on Instant Loan Platforms.

### 5.1 Perceived Benefits of Instant Loan Platforms

Participants who were successful in availing instant loans through the applications expressed great excitement and gratitude towards these platforms. While many of our participants faced significant financial hardships even before the COVID-19 pandemic, almost all of them experienced exacerbated difficulties during the pandemic. Several participants (n=16) reported seeking loans to either supplement or substitute their loss of incomes. These platforms also offered attractive benefits, giving our participants the perceptions of loans with no-strings attached. We highlight perceived benefits of instant loan platforms with the help of the following codes: 1) Being able to access anytime, anywhere, 2) Ensuring dignity and privacy, 3) Preserving social ties, and 4) Promising social mobility.

---
[3]credit scores generated by the Credit Information Bureau India Limited

*5.1.1 Being able to access money anytime, anywhere.* Participants' enthusiasm for instant loans often highlighted their distrust in formal banking sectors, a finding also reported in research on financial experiences of other vulnerable populations [95, 118]. Our participants despised extensive verification processes of formal loans. Formal loan processes required applicants to submit a long list of identity verification documents such as birth certificates, caste certificates, assets documents, employment certificates. Several of our participants did not have these documents to begin with,[4] shutting them out of formal financial systems. Finding the right set of documents to produce is never an easy task for anyone, and was exceptionally difficult for those participants who had lower levels of education and literacy. Further, participants were required to seek willing guarantors who would support their applications, open up their homes to unannounced visits from loan officers, and haggle with them for weeks, even after which there was no guarantee of a loan. As P16 said, *"[Banks] have several rules and regulations. [..] They might say, 'there was a server problem, come back tomorrow.' They make us roam everywhere. There are a lot of internal things that we don't understand."* In contrast, instant loans arrived into users' bank accounts within minutes of them making requests. While the requirements of different apps varied slightly, participants generally recalled providing minimal details such as the their names, addresses, phone numbers, selfies, permanent account numbers (PAN) and Aadhar card numbers (unique identification numbers). Given the convenience of loan apps, participants anecdotally mentioned borrowing loans from quaint locations such as under the streetlights and bathrooms at midnight. Such instant money was a *boon* for participants like P10 during dire circumstances: *"[W]hen I installed the app, the first thing I was happy about - instant cash. Because it was the urge of the money at that point of time. And I got within 5 minutes! Believe it or not 5 minutes, I received the money."*

*5.1.2 Ensuring dignity and privacy.* Several participants who praised the features of instant loans shared their experiences with local money lenders or pawn brokers, whom they had turned to due to difficulties in getting loans from banks. However, local lenders had often charged high interest rates, demanded repayment on their whims, and had sometimes employed aggressive recovery tactics such as visiting homes, and harassing our participants and their families. P3 recalled, *"I had taken once a 2000 rupees (USD 27) loan [from a local lender], and they had charged me 500 interest that I was supposed to repay within a month. They were harassing me, and wanted the money immediately."* Fearing reputation harm from local lenders and gossip in their social circles, participants avoided seeking such loans except in unavoidable circumstances. Instant loan applications, by nature of being on users' mobile devices, offered a high degree of privacy that was previously unavailable for participants. P19 explained, *"You take from market [...] everyone will get to know. [People] will talk. Here [...] no one else will know."* Users were afforded the flexibility of starting repayments after a few months, and could even request time extensions with 1-click features. Our participants welcomed such features as hallmarks of borrower-friendliness. P3, who was quoted previously contrasted their experiences, *"Whether your loan amount is large or small, [instant loan platforms] will give you some time to repay... [When local lenders were harassing me], I took this instant loan. [The app] gave me three months time to repay the money and interest was also just 300 rupees."* In addition, the promise of digital transactions instilled hopes of digital repercussions on defaulting, like a meagre impact on participants' CIBIL credit scores, thus increasing their overall comfort in borrowing instant loans.

*5.1.3 Preserving social ties.* Almost all our participants reported routinely turning to their closest circles during times of need. However, such lending-and-borrowing was riddled with complexities. First, it was difficult for participants to even muster the courage to ask their social circles. In social circles, small loans were indicative of participants' inability to manage their households, and thus hurt their respectability. When P14 asked relatives for help with her child's education, she received unsolicited advice in the guise of care: *"[They said], 'why do you want to send [your kid] to that school paying high fees in this critical situation? You can just shift [switch] to government (public) school."* Public schools in India offer free education, and are often viewed as schools for children from lower socio-economic backgrounds. Education is highly regarded as a mobility tool for the middle classes in India. Hence, sending children to well-regarded private schools is both a responsibility, and a matter for pride for parents, leading P14 to perceive the advice as derogatory. Given such humiliating experiences, participants equated borrowing from social circles with pledging their *"self-respect."* Naturally, when their requests for money were unmet, participants like P9 dealt with extreme feelings of rejection that strained relationships: *"I knew they had the money, and they still refused. That is why I don't feel like asking anyone money... Earlier I used to keep in touch regularly, now that bonding is not there."* Small, predetermined loans offered by instant loan platforms removed burdens of ask, and alleviated worries of social image for participants. In addition, participants hoped that the 'instant' nature of loans would ensure that they did not dwell on their feelings in case of rejection. As P26 said, *"If I get [the loan] I'm lucky enough. If not, [...] [t]here's no risk involved. [...] [Next day], there is a tendency that you could forget also. You would just move off (on)."*

*5.1.4 Promising social mobility.* Some participants had been lured into the credit system previously through shopping and entertainment, but such experiences had rarely ended well. They had subscribed to comforts and generous credit limit increases without worrying about monthly EMI payments. Few users had understood how credit cards worked, what credit scores meant, and the implications of defaulting on their credit bills. P20 learned about the implications of credit scoring system through negative experiences: *"First we got INR 5000 from [financial institution], which gradually increased to INR 15,000 [and finally reached] INR 1,60,000. I purchased many products through [credit][...]. I now have [a credit score of] 600. Because of this, nobody is giving us loans."* Participants with middle class aspirations were often fearful, and expressed aversion to incurring large debts. However, as P08 put it, seemingly small loans offered by instant loan platforms were necessary evils that could help users build credit and achieve dreams of mobility: *"[B]anks should grant us loans in the future. [...] If we don't take a loan, [credit score] will go in the negative."* Participants thus reported seeking

---
[4] Having access to birth certificates and caste certificates is highly correlated with class, caste and socio-economic status in India. As of 2016, 62.3% of children under the age of 5 had birth certificates, and 69.1% of all household members had aadhar cards [65]

instant loans for their children's education, for upgrading their comforts, and to secure financial independence.

## 5.2 Perceived Obligations to Instant Loan Platforms

Instant loan platforms were sources of immediate money and also the only means of survival for participants during difficult times when they ran from pillars to posts to seek financial help. Participants used instant loans to manage their everyday expenses, ranging from buying groceries, paying for school fees of their children, to clearing outstanding debts. Thus, as in the old adage, several participants equated the loan platforms with friends, and expressed intense feelings of indebtedness towards loan companies. P01, who sought instant loans when his business went haywire said, *"It really helped me during my tough times, so I actually owe them and I'm actually [still] owing them... I would recommend this app to so many of my contacts and I would say just like how 'a friend is in need is a friend indeed'."* Our participants perceived and fulfilled several obligations toward the loan platforms, that we explain through the following codes: a) Accepting harsh terms and conditions, b) Over-sharing sensitive data, and c) Making high fee payments.

*5.2.1 Accepting harsh terms and conditions.* In the anticipation of 'instant' money, several participants acknowledged that they had simply clicked on 'I agree' to terms and conditions of the apps, without expending the slightest effort to understand what they were consenting to. Likewise, very few participants recalled the specific terms that had been imposed by the apps, which we found ranged from peculiar to extremely harsh. For instance, P4 explained how he generated 3D views of his face instructed by a facial recognition bot, *"It will ask for a selfie. Turn both the sides, open the mouth... blink your eyes, rotate your head."* Obliging such requests were mandatory if participants wished to proceed with their applications. Some others recalled agreeing to potential legal actions and home visits in the case of defaulting on small loans. Such acceptances were viewed as mere formalities in getting access to progressively large credit limits. For instance, P3 recalled their speedy acceptance of terms, and the subsequent ascent in credit limits, *"[I]f you're not repaying the loan then they will take the legal action. They can also come to the home, and you have to pay a penalty of eight rupees per day... [Y]ou have to say okay to all these things. [...] After this, they will give you 500 (USD 7) rupees first. Once you repay, they will give you 1000 (USD 15) (and so on)."* Quite naturally, our participants did not expect to negotiate the terms and conditions of instant loans. In fact, they believed that if they were being offered money during a financially difficult time, they had an obligation to accept all the terms and conditions associated with the money. In addition, almost all our participants strongly believed that regardless of the terms and conditions, a loan once sought must be rightfully returned to its owner to 'restore justice.' Consequently, our participants assumed all responsibility for a loan borrowed, and frequently associated defaulting on loans with ideas of 'cheating' and 'injustice' to the lender. These ideas were also supported by participants' cultural and religious beliefs. P15 explained: *"[I]f you intend to cheat somebody, you shouldn't take a loan... If we have taken a loan, we must repay it correctly [...] [Else], we will face sazaa (punishment) from Allah."*

*5.2.2 Over-sharing sensitive data.* Several loan platforms also required access to users' media and gallery, phone books, Whatsapp and Gmail contacts, location information, financial transaction texts, app usage analytics, and other device information. We found that our participants perceived sharing such data as tests of credibility. For them, withholding data meant a lack of confidence in their own abilities to repay loans. Attempting to borrow loans despite not being confident was equivalent to 'cheating' loan platforms. P16 explained, *"It is okay if they collect information. If I have an intention to cheat then I should be scared. [...] If I am willing to repay fairly, I need not be scared."* We also found that participants' mental models of instant loans shaped their data-sharing practices in complex ways. Several participants expressed some discomfort sharing such data. Some associated sensitive data with ideas of 'intimacy.' As P15 put it, *"If they are tracking where I'm going and what I'm doing, it's like sharing my family background (colloquial: wife's background) with them."* Others discussed fears of misuse and online scams. Yet, all participants had either already enabled permissions unknowingly, or showed willingness to do so. P24 weighed her discomfort against the need for money and arrived at a compromise, *"I got a thought that they will hack. But at that time money was important... [N]ow in home loan we pledge papers, in gold loan we pledge gold, in the same way digitally we have to pledge all our information."* Being used to models of lending and borrowing where trust in the exchange was facilitated through the value of pledged assets, our participants 'pledged' sensitive data as high-credibility collateral assets.

*5.2.3 Making high fee payments.* Instant loans came at high initial costs to participants. Platforms charged processing fees, disbursal fees, down-payments, often taking away 20-25% of loan amounts during disbursal. This was in addition to the high floating interest rates (15 - 35%) and penalty charged by platforms. For P10, these high fees were small costs of the convenience during what were difficult times for her: *"in case we don't pay consecutively for a month, some charges are there, but I wouldn't call it a disadvantage. When you are getting all these advantages that's a common thing. That's perfectly fine."* In addition, participants discussed how repeatedly borrowing through the same platform easily offset such costs. They received attractive benefits like promotional codes, discounts, and better terms on new loans as rewards for their loyalty; these reciprocal exchanges were perceived as mutually beneficial by participants. Thus, several participants developed emotional attachments to loan platforms. P01's emotional attachment nudged him towards safeguarding the interests of loan platforms through high fees, *"I wouldn't recommend this app to a person who doesn't have intention to pay [high fees]... That's my, a little bit emotional attachment. [...] That's how the [platform] will be able to give salary to their associates and the people supporting them."*

## 5.3 Dependence on Instant Loan Platforms

FinTech companies often tout narratives of financial inclusion for 'unbanked' users through instant loans [5–7]. We found that such inclusion came at the cost of participants' dependence on loan platforms. Participants circumvented barriers to access loans, borrowed cyclically through loan platforms, tolerated abuse from predatory lenders, and shared responsibility for their negative experiences, potentially leading to their financial and technology exclusion. We discuss these findings with the following codes: 1) Circumventing algorithmic discrimination, 2) Recurring debts, 3) Tolerating abuse, and 4) Assuming responsibility for loan platforms' failures.

*5.3.1 Circumventing algorithmic discrimination.* While instant loan apps are designed as single-user applications, we found evidence of intermediated use among our participants, as is commonly reported in previous research on technology-use in the Global South [12, 41, 72, 110]. Participants sought the help of others, often immediate family members or trusted close friends, to download and navigate the apps, submit their applications, and manage payments. In some cases, we also found that participants attempted to borrow loans through others' devices and profiles. For instance, P10 borrowed through her husband's phone after intuitively recognizing that instant loan limits could be influenced by gender pay gap, and gendered patterns of digital activities. She even suggested a friend to borrow through her husband's phone saying, *"Maybe our salary cycle is less and husbands' salary cycles are more. And they have the credit cards and stuff. Maybe it's interlinked."* While such stereotype reinforcement through ADS could be viewed as potential barriers to access, evoking strong reactions in the West [22], our participants did not perceive them so. In fact, they underscored the importance of reading *intentions* in attributing experiences to discrimination. P07 acknowledged that disparate treatment could lead to unfair outcomes, but asserted that instant loan platforms did not intentionally discriminate based on gender: *"If they are giving less to women and more to men, it will not be correct. [...] But when it comes to loans, mostly they will give equal amounts to everybody."* Our participants shared similar views towards other issues of algorithmic discrimination. P08, who identified as dark-skinned, talked about whitening their face digitally to get around potential intersectional accuracy disparities [26] in instant loan technology: *"We could use 'FaceApp' to modify our looks. If they (loan platforms) are grand thieves, we are petty thieves. That is the only difference."*

*5.3.2 Recurring debts.* Instant loan platforms made borrowing money pleasurable for participants by offering gamified engagement. In addition to in-app discounts, surprise offers and virtual coins that we discussed earlier, we found that some apps also gamified credit limit increases; the platforms would first offer small amounts like INR 5,000-10,000 (USD 75-135); users would then 'unlock' higher credit limits when they neared their repayment terms, mimicking level-increases in virtual games. Some participants suspected that such gamified mechanisms were recovery nudges in disguise. P7 with an outstanding loan of INR 10,000 (USD 135) noted, *"Now that INR 2,00,000 (USD 2700) lock has opened up... I feel they opened the lock to show me that I will be eligible for a larger amount when I close this loan."* We found that these mechanisms, in addition to ready acceptances of instant loans by our participants, had led several of them to borrow beyond their capacity. Such participants had then engaged in cyclical borrowing from several different apps to *'balance'* their loans. P23 had once gotten into an addictive rhythm of unlocking higher credit limits in the loan apps: *"[Let's say] we have cleared the first level, so it seems like they have confidence in us, and have automatically increased the limit. [...] I didn't realize it then, and would end up accepting the loan in a hurry. [...] I would run here and there and borrow from friends to repay the loan. [...] I would take a loan from another app to repay the friend. I had loans from 4 apps at one point."* Other participants didn't consider themselves 'addicted' to instant loans, but regretted cyclical borrowing. They justified recurring loans as unavoidable by-products of their financial vulnerability and social obligations.

*5.3.3 Tolerating abuse.* We also found evidence of abuse in our study. Through loan platforms, some participants fell prey to predatory lenders who employed aggressive recovery tactics for small amounts of money, as little as INR 2000 (USD 27). Their tactics included repeatedly harassing borrowers for repayments through calls and texts, issuing threats of legal action, broadcasting sensitive information to borrowers' contacts on WhatsApp and other social media, shaming defaulters, targeted harassment of borrowers' contacts, and home visits. Digital medium allowed predatory lenders to abuse borrowers at scale. For instance, lenders performed semantic association on borrowers' contact lists to identify their close contacts (sometimes inaccurately) and harass them. Such tactics caused immense emotional and reputation harm to participants, and damaged their dignity. P4 encountered stigmatization in their social circles: *"They contacted my friends and family through WhatsApp. They shared my photo and published my details saying I had taken loan and hadn't repaid and started harassing them... Because of this I lost a lot of friends. I even had troubles with relatives. I ended up losing my job. [...] I was very upset but did not share it with anyone. At one point, I even tried to commit suicide."* Unfortunately, until December 2020, instant loan platforms had received little attention from the Reserve Bank of India (RBI). Thus, several predatory loan platforms had flourished, causing many borrowers to die by suicide [18]. Our participants were aware of such risks; yet, very few were critical of the platforms. Participants' emotional attachments made it challenging for them to seek accountability from the platforms. P08 described the extent of vulnerability, *"When we are unable to borrow from our friends, this loan app is helpful to us, just as a friend in need. When we don't even know where we can get money, this app decides automatically and at least grants us INR 1000 (USD 14). It does not see caste or religion or skin color. They simply believe us based on what we type (our data) [...]. So, we cannot find fault with the app. [...] This app helps us when all others have abandoned us."*

*5.3.4 Assuming responsibility for loan platforms' failures.* Our participants often viewed their negative experiences through the lenses of 'incompetency', and assigned self-blame for their experiences. Even if loan platforms were at fault, they were seemingly offering loans with no asset requirements; therefore, any rule or tactic was justifiable. P23, a survivor of abuse from a predatory loan platform reflected on their learnings: *"I was very firm about not availing app loans but since my friend suggested it I took it. [...] I did not think about whether we could repay the loan during difficult times. Corona has taught me a very good lesson."* Other negative experiences included losing money to fake apps, or being rejected by loan platforms without due explanations. Contrary to normative expectations of recourse, such negative technological experiences induced feelings of 'shame' in our participants who were less likely to share such experiences with their peers or seek help. In addition, participants' ardent optimism in technology and a lack of confidence in their technical abilities often led them to assume unfair responsibility for their negative experiences. P22 who was confident about her creditworthiness blamed her lack of technical skills for an unexplained loan rejection: *"Maybe I made some mistake while typing. Because if they look at my PAN card, they will definitely give me a loan. So I feel that there must be some kind of mistake that I made."* Unfortunately, for participants with futuristic outlook on technology, negative experiences reinforced their beliefs that they would never be the intended audience for 'high tech' applications, resulting in technology abandonment. As P2 put it, the doors to an AI-powered future remained closed to them: *"I felt that this gate had closed for me. I felt I shouldn't go around and ask for money, or on these apps."*

## 6 DISCUSSION

Our work fills a critical gap in the research on algorithmic accountability: we provide an understanding of social conditions of accountability through the experiences of (potentially) vulnerable users who are constrained in their capacity to seek accountability from technology providers. We situate these findings in the larger discourse on algorithmic accountability, and provide some suggestions for contextualizing the design of accountability mechanisms. We conclude with implications of our work for the use of alternative data in FinTech applications in the Global South.

### 6.1 Examining Power Relations in Algorithmic Accountability

Current discourse on algorithmic accountability rests on the existence of accountability relationships between technology providers responsible for causing harm through ADS, and the individuals experiencing harm through ADS (or their representatives) [121]. In this relationship, the technology providers are *obligated* to provide 'accounts' to the those individuals who are harmed [15, 96, 101, 123]; these individuals or their representatives may then judge the accounts and seek to impose consequences if necessary. Consequently, much work in algorithmic accountability often presents 'sharing of information' by technology providers as the first phase of accountability [53, 64, 89, 121]. Prior work calls for involving affected individuals in designing accountability mechanisms to ensure that the information is meaningful to them [42, 75]. Our work extends this argument to show that purely technical approaches to accountability obscure the socio-political realities of stakeholders that make such 'information sharing' necessary in the first place.

In our study, exchanges enabled by AI-based instant loans reconfigured users' relations to instant loan platforms in ways that distract from the goals of algorithmic accountability. First, users were placed into positions of 'indebtedness' with loan platforms. Users in our study were largely 'thin-file' borrowers, making it difficult for them to secure loans from formal financial institutions. They had primarily relied on informal loans for their borrowing needs, which had come with huge social costs to them. Thus, instant loans, with seemingly no-collateral-requirement, no-strings-attached were viewed as a huge 'favor' by users. Under users' *debt relationships* with loan platforms, it was the users, rather than the platforms, who perceived obligations. Users fulfilled these obligations in both material and intangible ways, and persisted despite human and other costs, such as abuse, discrimination, recurring debts, privacy harms, and self-harm to them. Contrary to the normative behaviors of outrage in users documented from work in the West [101], users in our study did not believe it was in their right to question the terms and conditions of lenders. Instead, they assumed responsibility for their failures of loan platforms, thus demonstrating a dependence, and releasing those high-powered actors from the obligations of accountability. Thus, we argue that algorithmic accountability is mediated through platform-user power relations, and can be stymied by on-the-ground socio-political conditions of users. Responsible development of AI cannot be universally achieved without paying close attention to these situated [114] power dynamics. We need more research on the relationship between accountability mechanisms, agency of users, and the impetus for action in different socio-political contexts to ensure responsible AI more widely. We build on the work of Katell et al.[74], and propose a situated approach to algorithmic accountability.[5]

*6.1.1 Enhancing agency of the forum through critical awareness.* New internet users, with vastly different mental models of AI can place misguided trust in ADS [82, 93, 109]. Such high user-trust in AI systems played out in several ways in our study: ready acceptances of terms, conditions, and loan decisions, often to the extent of users reevaluating their own competencies and abilities. However, design and research in user-centered AI often assumes low trust in AI, and begins with questions of 'how might we design for *increased* user trust in AI'? Instead, designs must plan for appropriate failures assuming high-user trust in AI systems [14]. Research must address questions such as *decreasing user trust* or *increasing user distrust* in AI systems. Further, we saw that users who benefited from the instant loan applications developed deep emotional attachment towards these applications. This suggests that users' mental models of AI systems must be calibrated appropriately and at regular intervals of use. On-boarding users to AI systems via guides may be a viable first step to align users' mental models with AI systems [29]. However, such measures must be complemented by widespread AI literacy programs. Trust and safety initiative for users in India by Google is one such example [66]. More support must be given to grassroots organizations that are working to raise public awareness. An outstanding example is Internet Freedom Foundation's Project Panoptic that is raising awareness on public-facing facial recognition systems in India [67]. Such efforts must be supported by programs that not only up-skill citizens to be AI designers, but also critical thinkers who can be AI testers and AI auditors. These initiatives can help recognize the largely invisible work of *maintenance and repair* involved in responsibly deploying AI [44].

*6.1.2 Enabling meaningful transparency through collective spaces.* Transparency is a widely called for mechanism for accountability [8]. Making registries of datasets, models and processes available for public scrutiny [16, 59, 99] is a good first step. However, lack of technical expertise among the public could render such transparency meaningless. Thus, corporate actors and governments must work with civil advocacy groups to create toolkits that consumer advocates can use towards accountability efforts. The Algorithmic Equity toolkit by ACLU Washington could serve as a model for such aims [74]. Further, for transparency to serve the goal of answerability, it must generate sufficient pressure from the forum that forces actors to respond to violations. When platform-user

---

[5]We use situated accountability differently from that of Henriksen et al., who refer to the need for situating accountability policies in practices of designers' and engineers' working on the development of AI systems [62]

relations are entrenched in power differences, individual actions by vulnerable users may not be successful in large scale social changes [39, 78]. Therefore, we must go beyond transparency for individual users, and towards transparency of collective users. One way to achieve this may be through designing spaces where vulnerable users can mobilize support towards demanding collective accountability. Through our study, we saw that new internet users are often ashamed of their negative experiences, making it unlikely for them to share their experiences with other users offline. Anonymity provided by digital ecosystems can be leveraged to reduce such barriers for them. Such a platform could also lead to normalization of negative experiences, leading to discourse and then political action. Ahmed et al's Protibadi [13], a system to mobilize support against sexual harassment in Bangaladesh, and Irani and Silberman's Turkopticon [68] to invert requester-turk worker power relations are examples of intervention opportunities for researchers interested in algorithmic accountability.

*6.1.3 Re-configuring designer-user relations through community engagement.* Algorithmic harms such as bias and discrimination are extensively studied in FAccT, and receive extensive attention especially in Western media [26, 61, 87, 123]. However, we saw in our study that users accessing instant loans were undeterred by algorithmic discrimination. Rather, they expressed significant concerns about alternate forms of harms from ADS systems such as data leaks, gossip in social circles from data leaks, reputation damage and social frictions. Prior work has already pointed to the need to re-contextualize harm measurements [109]. We extend this argument and draw on work by Metcalf et al. to suggest that we must *co-construct measurements of harms* with the community of stakeholders involved [88]. While doing so, we must also recognize that a purely computation framing of harms fails to address injustices caused by structural oppression [63, 77]. Such structural oppression is at the root of what has 'excluded' these individuals from technology spaces, and created designer-user binaries. We therefore echo the calls made by scholars to reconfigure these relations through design practices situated in community values [33]. Design Beku [104] is an excellent model for how this could be done.

*6.1.4 Committing to justice through critical self-reflection.* Users behaviors towards AI-based predatory applications including justification, tolerance, acceptance and self-blame led to extreme consequences such as abuse, reputation harm and self-harm. Such experiences are a violation of users' privacy, and users' right to dignity. Thus, the findings in our study also point that responsible AI is a human rights' issue. What recourse mechanisms can we afford to these users in the case of undergoing data leaks that are the equivalent of emotional harm? Further, what recourse mechanisms can we afford to these users when there is *intentional* reputation harm? How far we can go in addressing human rights' issues with technical interventions? What does accountability mean when predatory lenders create mobile applications with open-sourced machine learning algorithms and datasets, and slap a usable interface to prey on vulnerable users? Can our radical vision of democratizing AI hurt more than help? What forms of accountability can be assumed when the tools we created land into the hands of malicious actors? While we acknowledge that we do not have the answers to these large challenges, we believe critical reflection might be a good first step.

## 6.2 Implications for the use of Alternative Data in FinTech Applications in India and Beyond

'Alternative lending' uses mobile phone data to solve information asymmetry problems of lenders, who traditionally depend on tangible collateral assets [57]. Such models could also carry huge benefits to borrowers: As we saw in our study, they could open up opportunities for users who have never been a part of formal financial systems. These benefits were especially significant to our participants given the economic challenges brought on by the COVID-19 pandemic. Unfortunately, alternative lending could also have extreme downsides; without regulation or rules to define the limits of what counts as 'alternative data', the judgements made based on these data are largely arbitrary. In addition, new internet users in the Global South (such as the users in our study), may overshare sensitive data in the name of high quality *collateral assets* to unverified platforms, risking privacy harms. Current techniques around privacy, data rights and data sovereignty rarely account for data as collateral assets, calling for research to re-frame designs around privacy, safety and trust. The harms of alternative credit often extend beyond the instance of decision-making. That is, data assets can themselves be *elite resources* [111], and are often the products of uneven social relations [38]. For instance, loan platforms reproduced gender relations prevalent in economic and social spheres when women participants used their husbands' phones to seek loans. If the goal of AI-based lending is to achieve equitable financial inclusion, we must account for such data disparities in our imaginations of AI systems. Further, data collection mechanisms may be *predatory*. Users in our study reported receiving ads on their phones even when they were unsuccessful with the apps, or several months after they had stopped using the apps. While one could argue that such predatory mechanisms could be curtailed with better user privacy, we remind the reader that giving consent and accepting privacy policies were unparalleled obligations to financially stressed users in comparison to 'instant' cash. New data privacy and consent models such as collective consent [107] may be viable options. As an immediate call-to-action, we urge designers to implement AI systems based on established industry practices [15, 96]. Such practices include sourcing data responsibly i.e., ensuring that users' personally identifying information is protected at all times, preparing a data-maintenance plan for the life-cycle of the product, collecting routine user feedback, aligning feedback with model improvements, and communicating the value and time-to-impact to users, identifying factors that go into user trust, helping users calibrate their trust, calibrating trust through the product experience, and managing influence on user decisions [15, 96, 111]. We also call on designers to supplement these efforts with awareness campaigns on data and privacy rights for vulnerable users. Beyond these implications, our work opens up policy questions such as: How do we communicate the potential risks of 'instant' money to users in dire circumstances? What educational and financial aid would they need? Who should assume responsibility? We believe these could be important future directions.

## 7 CONCLUSION

We presented a qualitative study of 29 financially-stressed users' interactions with instant loan platforms in India. We reported on the perceptions of instant loan platforms among users, and their feelings of 'indebtedness' towards those platforms. We elaborated on the ways in which these users fulfilled obligations, and enacted dependence on loan platforms. By situating our findings in the algorithmic accountability discourse, we presented an argument that algorithmic accountability is mediated through platform-user power relations, and can be hindered by on-the-ground socio-political conditions of users. We proposed situated accountability interventions such as enhancing agency of the forum, enabling collective transparency, reconfiguring designer-user relations, and committing to critical self-reflection to ensure wider accountability. We conclude with implications for FinTech applications in India and beyond.

## ACKNOWLEDGMENTS

We thank Azhagu Meena S P for assisting with interviews, and Vinodkumar Prabhakaran, Nikola Banovic, Jane Im, Nel Escher and Anindya Das Antar for helpful feedback on this work. We also thank the reviewers at CHI'22 where a previous draft was first submitted, and the reviewers of FAccT for their helpful comments. Finally, we thank our participants who trusted us with their experiences; without them this research would have never been possible.

## REFERENCES

[1] 2021. Non-Banking Financial Company. https://www.rbi.org.in/Scripts/FAQView.aspx?Id=92
[2] 2022. Dhani - India's Trusted Site for Finance, Healthcare and Online Medicines. https://www.dhani.com/
[3] 2022. Five ways that AI augments FinTech. https://indiaai.gov.in/article/five-ways-that-ai-augments-fintech
[4] 2022. Get line of credit up to Rs. 5 Lakhs - MoneyTap. https://www.moneytap.com
[5] 2022. Kissht. https://kissht.com
[6] 2022. KreditBee. https://www.kreditbee.in/
[7] 2022. SmartCoin - Get Instant Credit. https://smartcoin.co.in/
[8] Ashraf Abdul, Jo Vermeulen, Danding Wang, Brian Y Lim, and Mohan Kankanhalli. 2018. Trends and trajectories for explainable, accountable and intelligible systems: An hci research agenda. In *Proceedings of the 2018 CHI conference on human factors in computing systems*. 1–18.
[9] Ada Lovelace Institute, AI Now Institute, and Open Government Partnership. 2021. Algorithmic Accountability for the Public Sector. (2021).
[10] Amina Adadi and Mohammed Berrada. 2018. Peeking inside the black-box: a survey on explainable artificial intelligence (XAI). *IEEE access* 6 (2018), 52138–52160.
[11] Sray Agarwal. 2020. AI powered FinTech: The drivers of Digital India. https://indiaai.gov.in/article/ai-powered-fintech-the-drivers-of-digital-india
[12] Syed Ishtiaque Ahmed, Md Romael Haque, Shion Guha, Md Rashidujjaman Rifat, and Nicola Dell. 2017. Privacy, security, and surveillance in the Global South: A study of biometric mobile SIM registration in Bangladesh. In *Proceedings of the 2017 CHI Conference on Human Factors in Computing Systems*. 906–918.
[13] Syed Ishtiaque Ahmed, Steven J Jackson, Nova Ahmed, Hasan Shahid Ferdous, Md Rashidujjaman Rifat, ASM Rizvi, Shamir Ahmed, and Rifat Sabbir Mansur. 2014. Protibadi: A platform for fighting sexual harassment in urban Bangladesh. In *Proceedings of the SIGCHI Conference on Human Factors in Computing Systems*. 2695–2704.
[14] Saleema Amershi. 2020. Toward Responsible AI by Planning to Fail. In *Proceedings of the 26th ACM SIGKDD International Conference on Knowledge Discovery & Data Mining*. 3607–3607.
[15] Saleema Amershi, Dan Weld, Mihaela Vorvoreanu, Adam Fourney, Besmira Nushi, Penny Collisson, Jina Suh, Shamsi Iqbal, Paul N Bennett, Kori Inkpen, et al. 2019. Guidelines for human-AI interaction. In *Proceedings of the 2019 chi conference on human factors in computing systems*. 1–13.
[16] Amsterdam. 2020. Algorithmic Register Amsterdam. https://algoritmeregister.amsterdam.nl/en/ai-register/
[17] Mike Ananny and Kate Crawford. 2018. Seeing without knowing: Limitations of the transparency ideal and its application to algorithmic accountability. *new media & society* 20, 3 (2018), 973–989.
[18] Varsha Bansal. 2021. Shame, suicide and the dodgy loan apps plaguing Google's Play Store. *Wired UK* (2021). https://www.wired.co.uk/article/google-loan-apps-india-deaths
[19] Alistair Barr. 2015. Google Mistakenly Tags Black People as 'Gorillas,' Showing Limits of Algorithms. *Wall Street Journal* (jul 2015). https://www.wsj.com/articles/BL-DGB-42522
[20] Ruha Benjamin. 2019. Race after technology: Abolitionist tools for the new jim code. *Social Forces* (2019).
[21] Reuben Binns, Max Van Kleek, Michael Veale, Ulrik Lyngs, Jun Zhao, and Nigel Shadbolt. 2018. 'It's Reducing a Human Being to a Percentage' Perceptions of Justice in Algorithmic Decisions. In *Proceedings of the 2018 Chi conference on human factors in computing systems*. 1–14.
[22] Tolga Bolukbasi, Kai-Wei Chang, James Y Zou, Venkatesh Saligrama, and Adam T Kalai. 2016. Man is to computer programmer as woman is to homemaker? debiasing word embeddings. *Advances in neural information processing systems* 29 (2016), 4349–4357.
[23] Danah Boyd and Kate Crawford. 2012. Critical questions for big data: Provocations for a cultural, technological, and scholarly phenomenon. *Information, communication & society* 15, 5 (2012), 662–679.
[24] Virginia Braun and Victoria Clarke. 2012. Thematic analysis. (2012).
[25] Anna Brown, Alexandra Chouldechova, Emily Putnam-Hornstein, Andrew Tobin, and Rhema Vaithianathan. 2019. Toward algorithmic accountability in public services: A qualitative study of affected community perspectives on algorithmic decision-making in child welfare services. In *Proceedings of the 2019 CHI Conference on Human Factors in Computing Systems*. 1–12.
[26] Joy Buolamwini and Timnit Gebru. 2018. Gender shades: Intersectional accuracy disparities in commercial gender classification. In *Conference on fairness, accountability and transparency*. PMLR, 77–91.
[27] Jenna Burrell. 2016. How the machine 'thinks': Understanding opacity in machine learning algorithms. *Big Data & Society* 3, 1 (2016), 2053951715622512.
[28] CAC. 2021. Internet Information Service Algorithm Recommendation Management Regulations. http://www.cac.gov.cn/2021-08/27/c_1631652502874117.htm.
[29] Carrie J Cai, Samantha Winter, David Steiner, Lauren Wilcox, and Michael Terry. 2019. " Hello AI": Uncovering the Onboarding Needs of Medical Practitioners for Human-AI Collaborative Decision-Making. *Proceedings of the ACM on Human-computer Interaction* 3, CSCW (2019), 1–24.
[30] Diogo V Carvalho, Eduardo M Pereira, and Jaime S Cardoso. 2019. Machine learning interpretability: A survey on methods and metrics. *Electronics* 8, 8 (2019), 832.
[31] Husanjot Chahal, Sara Abdulla, Jonathan Murdick, and Ilya Rahkovsky. 2021. *Mapping India's AI Potential*. Technical Report.
[32] C Chausson. 2016. France opens the source code of tax and benefits calculators to increase transparency.
[33] Sasha Costanza-Chock. 2020. *Design justice: Community-led practices to build the worlds we need*.
[34] Aman Dalmia, Jerome White, Ankit Chaurasia, Vishal Agarwal, Rajesh Jain, Dhruvin Vora, Balasaheb Dhame, Raghu Dharmaraju, and Rahul Panicker. 2020. Pest Management In Cotton Farms: An AI-System Case Study from the Global South. In *Proceedings of the 26th ACM SIGKDD International Conference on Knowledge Discovery & Data Mining*. 3119–3127.
[35] Nicholas Diakopoulos. 2015. Algorithmic accountability: Journalistic investigation of computational power structures. *Digital journalism* 3, 3 (2015), 398–415.
[36] Nicholas Diakopoulos. 2017. Enabling accountability of algorithmic media: transparency as a constructive and critical lens. In *Transparent data mining for Big and Small Data*. Springer, 25–43.
[37] Nicholas Diakopoulos and Michael Koliska. 2017. Algorithmic transparency in the news media. *Digital journalism* 5, 7 (2017), 809–828.
[38] Catherine D'ignazio and Lauren F Klein. 2020. *Data feminism*. MIT press.
[39] Paul Dourish. 2010. HCI and environmental sustainability: the politics of design and the design of politics. In *Proceedings of the 8th ACM conference on designing interactive systems*. 1–10.
[40] Julia Dressel and Hany Farid. 2018. The accuracy, fairness, and limits of predicting recidivism. *Science advances* (2018).
[41] Michaelanne Dye, David Nemer, Laura R Pina, Nithya Sambasivan, Amy S Bruckman, and Neha Kumar. 2017. Locating the Internet in the Parks of Havana. In *Proceedings of the 2017 CHI Conference on Human Factors in Computing Systems*. 3867–3878.
[42] Upol Ehsan, Q Vera Liao, Michael Muller, Mark O Riedl, and Justin D Weisz. 2021. Expanding explainability: Towards social transparency in ai systems. In *Proceedings of the 2021 CHI Conference on Human Factors in Computing Systems*. 1–19.
[43] Upol Ehsan, Samir Passi, Q Vera Liao, Larry Chan, I Lee, Michael Muller, Mark O Riedl, et al. 2021. The Who in Explainable AI: How AI Background Shapes Perceptions of AI Explanations. *arXiv preprint arXiv:2107.13509* (2021).

[44] MC Elish and EA Watkins. 2020. Repairing innovation: A study of integrating AI in clinical care. *Unpublished Manuscript* (2020).
[45] Isil Erel, Lea H Stern, Chenhao Tan, and Michael S Weisbach. 2018. Could machine learning help companies select better board directors? *Harvard Business Review* 1, 5 (2018).
[46] Nel Escher and Nikola Banovic. 2020. Exposing Error in Poverty Management Technology: A Method for Auditing Government Benefits Screening Tools. *Proceedings of the ACM on Human-Computer Interaction* 4, CSCW1 (2020), 1–20.
[47] Motahhare Eslami, Aimee Rickman, Kristen Vaccaro, Amirhossein Aleyasen, Andy Vuong, Karrie Karahalios, Kevin Hamilton, and Christian Sandvig. 2015. " I always assumed that I wasn't really that close to [her]" Reasoning about Invisible Algorithms in News Feeds. In *Proceedings of the 33rd annual ACM conference on human factors in computing systems*. 153–162.
[48] Motahhare Eslami, Kristen Vaccaro, Min Kyung Lee, Amit Elazari Bar On, Eric Gilbert, and Karrie Karahalios. 2019. User attitudes towards algorithmic opacity and transparency in online reviewing platforms. In *Proceedings of the 2019 CHI Conference on Human Factors in Computing Systems*. 1–14.
[49] ET Goverment. 2021. Odisha launches AI based online life certificate system for pensioners.
[50] Virginia Eubanks. 2018. *Automating inequality: How high-tech tools profile, police, and punish the poor*. St. Martin's Press.
[51] Simson Garfinkel, Jeanna Matthews, Stuart S Shapiro, and Jonathan M Smith. 2017. Toward algorithmic transparency and accountability.
[52] Susan Wharton Gates, Vanessa Gail Perry, and Peter M Zorn. 2002. Automated underwriting in mortgage lending: Good news for the underserved? *Housing Policy Debate* 13, 2 (2002), 369–391.
[53] Timnit Gebru, Jamie Morgenstern, Briana Vecchione, Jennifer Wortman Vaughan, Hanna Wallach, Hal Daumé Iii, and Kate Crawford. 2021. Datasheets for datasets. *Commun. ACM* 64, 12 (2021), 86–92.
[54] Tarleton Gillespie. 2014. The relevance of algorithms. *Media technologies: Essays on communication, materiality, and society* 167, 2014 (2014), 167.
[55] Leilani H Gilpin, David Bau, Ben Z Yuan, Ayesha Bajwa, Michael Specter, and Lalana Kagal. 2018. Explaining explanations: An overview of interpretability of machine learning. In *2018 IEEE 5th International Conference on data science and advanced analytics (DSAA)*. IEEE, 80–89.
[56] Government of Tamil Nadu. 2020. *Tamil Nadu Safe and Ethical Artificial Intelligence Policy*. Technical Report.
[57] Darrell Grissen et al. 2019. *Behavior Revealed in Mobile Phone Usage Predicts Loan Repayment*. Technical Report. arXiv. org.
[58] Varun Gulshan, Lily Peng, Marc Coram, Martin C Stumpe, Derek Wu, Arunachalam Narayanaswamy, Subhashini Venugopalan, Kasumi Widner, Tom Madams, Jorge Cuadros, et al. 2016. Development and validation of a deep learning algorithm for detection of diabetic retinopathy in retinal fundus photographs. *Jama* 316, 22 (2016), 2402–2410.
[59] M Haataja, L van de Fliert, and P Rautio. 2020. Public AI Registers: Realising AI transparency and civic participation in government use of AI. (2020).
[60] Alexa Hagerty and Igor Rubinov. 2019. Global AI ethics: a review of the social impacts and ethical implications of artificial intelligence. *arXiv preprint arXiv:1907.07892* (2019).
[61] Sara Hajian, Francesco Bonchi, and Carlos Castillo. 2016. Algorithmic bias: From discrimination discovery to fairness-aware data mining. In *Proceedings of the 22nd ACM SIGKDD international conference on knowledge discovery and data mining*. 2125–2126.
[62] Anne Henriksen, Simon Enni, and Anja Bechmann. 2021. Situated accountability: Ethical principles, certification standards, and explanation methods in applied AI. In *Proceedings of the 2021 AAAI/ACM Conference on AI, Ethics, and Society*. 574–585.
[63] Anna Lauren Hoffmann. 2019. Where fairness fails: data, algorithms, and the limits of antidiscrimination discourse. *Information, Communication & Society* 22, 7 (2019), 900–915.
[64] Ben Hutchinson, Andrew Smart, Alex Hanna, Emily Denton, Christina Greer, Oddur Kjartansson, Parker Barnes, and Margaret Mitchell. 2021. Towards accountability for machine learning datasets: Practices from software engineering and infrastructure. In *Proceedings of the 2021 ACM Conference on Fairness, Accountability, and Transparency*. 560–575.
[65] ICF IIPS. 2017. India National Family Health Survey NFHS-4 2015–16. *Mumbai: IIPS and ICF* (2017).
[66] Google India. 2022. Helping users stay safe online. https://forindia.withgoogle.com/intl/en/trust-and-safety/
[67] Internet Freedom Foundation. [n.d.]. Project Panoptic: Facial Recognition Systems in India.
[68] Lilly C Irani and M Six Silberman. 2013. Turkopticon: Interrupting worker invisibility in amazon mechanical turk. In *Proceedings of the SIGCHI conference on human factors in computing systems*. 611–620.
[69] Mohammad Hossein Jarrahi, Gemma Newlands, Min Kyung Lee, Christine T Wolf, Eliscia Kinder, and Will Sutherland. 2021. Algorithmic management in a work context. *Big Data & Society* 8, 2 (2021), 20539517211020332.
[70] Maria K, Styliani Kleanthous, Pınar Barlas, and Jahna Otterbacher. 2021. I Agree with the Decision, but They Didn't Deserve This: Future Developers' Perception of Fairness in Algorithmic Decisions. In *Proceedings of the 2021 ACM Conference on Fairness, Accountability, and Transparency* (Virtual Event, Canada) *(FAccT '21)*. Association for Computing Machinery, New York, NY, USA, 690–700. https://doi.org/10.1145/3442188.3445931
[71] Shivaram Kalyanakrishnan, Rahul Alex Panicker, Sarayu Natarajan, and Shreya Rao. 2018. Opportunities and challenges for artificial intelligence in India. In *Proceedings of the 2018 AAAI/ACM conference on AI, Ethics, and Society*. 164–170.
[72] Vaishnav Kameswaran and Srihari Hulikal Muralidhar. 2019. Cash, Digital Payments and Accessibility: A Case Study from Metropolitan India. *Proceedings of the ACM on Human-Computer Interaction* 3, CSCW (2019), 1–23.
[73] Shivani Kapania, Oliver Siy, Gabe Clapper, Azhagu SP, and Nithya Sambasivan. 2022. "Because AI is 100% right and safe": User Attitudes and Sources of AI Authority in India. In *CHI Conference on Human Factors in Computing Systems (CHI '22)*.
[74] Michael Katell, Meg Young, Dharma Dailey, Bernease Herman, Vivian Guetler, Aaron Tam, Corinne Bintz, Daniella Raz, and PM Krafft. 2020. Toward situated interventions for algorithmic equity: lessons from the field. In *Proceedings of the 2020 conference on fairness, accountability, and transparency*. 45–55.
[75] Harmanpreet Kaur, Harsha Nori, Samuel Jenkins, Rich Caruana, Hanna Wallach, and Jennifer Wortman Vaughan. 2020. Interpreting Interpretability: Understanding Data Scientists' Use of Interpretability Tools for Machine Learning. In *Proceedings of the 2020 CHI Conference on Human Factors in Computing Systems*. 1–14.
[76] Jakko Kemper and Daan Kolkman. 2019. Transparent to whom? No algorithmic accountability without a critical audience. *Information, Communication & Society* 22, 14 (2019), 2081–2096.
[77] Os Keyes, Jevan Hutson, and Meredith Durbin. 2019. A mulching proposal: Analysing and improving an algorithmic system for turning the elderly into high-nutrient slurry. In *Extended Abstracts of the 2019 CHI Conference on Human Factors in Computing Systems*. 1–11.
[78] Bran Knowles, Lynne Blair, Mike Hazas, and Stuart Walker. 2013. Exploring sustainability research in computing: where we are and where we go next. In *Proceedings of the 2013 ACM international joint conference on Pervasive and ubiquitous computing*. 305–314.
[79] Steinar Kvale. 2008. *Doing interviews*. Sage.
[80] Ms Amina Lahreche, Ms Sumiko Ogawa, Ms Kimberly Beaton, Purva Khera, Majid Bazarbash, Mr Ulric Eriksson von Allmen, Ms Ratna Sahay, et al. 2020. *The Promise of Fintech: Financial Inclusion in the Post COVID-19 Era*. Technical Report. International Monetary Fund.
[81] Min Kyung Lee, Daniel Kusbit, Evan Metsky, and Laura Dabbish. 2015. Working with machines: The impact of algorithmic and data-driven management on human workers. In *Proceedings of the 33rd annual ACM conference on human factors in computing systems*. 1603–1612.
[82] Min Kyung Lee and Katherine Rich. 2021. Who Is Included in Human Perceptions of AI?: Trust and Perceived Fairness around Healthcare AI and Cultural Mistrust. In *Proceedings of the 2021 CHI Conference on Human Factors in Computing Systems*. 1–14.
[83] Q Vera Liao, Daniel Gruen, and Sarah Miller. 2020. Questioning the AI: informing design practices for explainable AI user experiences. In *Proceedings of the 2020 CHI Conference on Human Factors in Computing Systems*. 1–15.
[84] Ryan Mac. 2021. Facebook Apologizes After A.I. Puts 'Primates' Label on Video of Black Men. *The New York Times* (sep 2021). https://www.nytimes.com/2021/09/03/technology/facebook-ai-race-primates.html
[85] Vidushi Marda. 2018. *Artificial Intelligency Policy in India: A Framework for Engaging the Limits of Data-Driven Decision Making*. Technical Report.
[86] Vidushi Marda and Shivangi Narayan. 2020. Data in New Delhi's predictive policing system. In *Proceedings of the 2020 conference on fairness, accountability, and transparency*. 317–324.
[87] Ninareh Mehrabi, Fred Morstatter, Nripsuta Saxena, Kristina Lerman, and Aram Galstyan. 2021. A survey on bias and fairness in machine learning. *ACM Computing Surveys (CSUR)* 54, 6 (2021), 1–35.
[88] Jacob Metcalf, Emanuel Moss, Elizabeth Anne Watkins, Ranjit Singh, and Madeleine Clare Elish. 2021. Algorithmic impact assessments and accountability: The co-construction of impacts. In *Proceedings of the 2021 ACM Conference on Fairness, Accountability, and Transparency*. 735–746.
[89] Margaret Mitchell, Simone Wu, Andrew Zaldivar, Parker Barnes, Lucy Vasserman, Ben Hutchinson, Elena Spitzer, Inioluwa Deborah Raji, and Timnit Gebru. 2019. Model cards for model reporting. In *Proceedings of the conference on fairness, accountability, and transparency*. 220–229.
[90] Emanuel Moss, Elizabeth Anne Watkins, Ranjit Singh, Madeleine Clare Elish, and Jacob Metcalf. 2021. Assembling Accountability: Algorithmic Impact Assessment for the Public Interest. *Available at SSRN 3877437* (2021).
[91] NITI Aayog. 2021. Responsible AI #AIFORALL. (2021).
[92] Safiya Umoja Noble. 2018. *Algorithms of oppression*. New York University Press.

[93] Chinasa T Okolo, Srujana Kamath, Nicola Dell, and Aditya Vashistha. 2021. "It cannot do all of my work": Community Health Worker Perceptions of AI-Enabled Mobile Health Applications in Rural India. In *Proceedings of the 2021 CHI Conference on Human Factors in Computing Systems*. 1–20.

[94] Cathy O'neil. 2016. *Weapons of math destruction: How big data increases inequality and threatens democracy*. Crown.

[95] Jacki O'neill, Anupama Dhareshwar, and Srihari H Muralidhar. 2017. Working digital money into a cash economy: The collaborative work of loan payment. *Computer Supported Cooperative Work (CSCW)* 26, 4 (2017), 733–768.

[96] Google PAIR. 2019. People + AI Guidebook. https://design.google/ai-guidebook

[97] Joyojeet Pal, Priyank Chandra, Vaishnav Kameswaran, Aakanksha Parameshwar, Sneha Joshi, and Aditya Johri. 2018. Digital payment and its discontents: Street shops and the Indian government's push for cashless transactions. In *Proceedings of the 2018 CHI Conference on Human Factors in Computing Systems*. 1–13.

[98] Frank Pasquale. 2015. *The black box society*. Harvard University Press.

[99] S Pénicaud. 2021. Building Public Algorithm Registers: Lessons Learned from the French Approach. (2021).

[100] Emilee Rader, Kelley Cotter, and Janghee Cho. 2018. Explanations as mechanisms for supporting algorithmic transparency. In *Proceedings of the 2018 CHI conference on human factors in computing systems*. 1–13.

[101] Inioluwa Deborah Raji and Joy Buolamwini. 2019. Actionable auditing: Investigating the impact of publicly naming biased performance results of commercial ai products. In *Proceedings of the 2019 AAAI/ACM Conference on AI, Ethics, and Society*. 429–435.

[102] Inioluwa Deborah Raji, Andrew Smart, Rebecca N White, Margaret Mitchell, Timnit Gebru, Ben Hutchinson, Jamila Smith-Loud, Daniel Theron, and Parker Barnes. 2020. Closing the AI accountability gap: Defining an end-to-end framework for internal algorithmic auditing. In *Proceedings of the 2020 conference on fairness, accountability, and transparency*. 33–44.

[103] Bogdana Rakova, Jingying Yang, Henriette Cramer, and Rumman Chowdhury. 2021. Where Responsible AI Meets Reality: Practitioner Perspectives on Enablers for Shifting Organizational Practices. *Proc. ACM Hum.-Comput. Interact.* 5, CSCW1, Article 7 (apr 2021), 23 pages. https://doi.org/10.1145/3449081

[104] Padmini Ray Murray, Naveen L Bagalkot, Shreyas Srivatsa, and Paul Anthony. 2021. Design Beku: Toward Decolonizing Design and Technology through Collaborative and Situated Care-in-Practices. *Global Perspectives* 2, 1 (2021), 26132.

[105] Marco Tulio Ribeiro, Sameer Singh, and Carlos Guestrin. 2016. Model-agnostic interpretability of machine learning. *arXiv preprint arXiv:1606.05386* (2016).

[106] Rashida Richardson. 2021. Defining and Demystifying Automated Decision Systems. *Maryland Law Review, Forthcoming* (2021).

[107] A Ruhaak. 2021. When One Affects Many: The Case For Collective Consent. *Mozilla Foundation* 20 (2021).

[108] Johnny Saldaña. 2015. *The coding manual for qualitative researchers*. Sage.

[109] Nithya Sambasivan, Erin Arnesen, Ben Hutchinson, Tulsee Doshi, and Vinodkumar Prabhakaran. 2021. Re-imagining algorithmic fairness in india and beyond. In *Proceedings of the 2021 ACM Conference on Fairness, Accountability, and Transparency*. 315–328.

[110] Nithya Sambasivan, Ed Cutrell, Kentaro Toyama, and Bonnie Nardi. 2010. Intermediated technology use in developing communities. In *Proceedings of the SIGCHI Conference on Human Factors in Computing Systems*. 2583–2592.

[111] Nithya Sambasivan, Shivani Kapania, Hannah Highfill, Diana Akrong, Praveen Paritosh, and Lora M Aroyo. 2021. "Everyone wants to do the model work, not the data work": Data Cascades in High-Stakes AI. In *proceedings of the 2021 CHI Conference on Human Factors in Computing Systems*. 1–15.

[112] Christian Sandvig, Kevin Hamilton, Karrie Karahalios, and Cedric Langbort. [n.d.]. Auditing algorithms: Research methods for detecting discrimination on internet platforms. ([n. d.]).

[113] Alison Smith-Renner, Ron Fan, Melissa Birchfield, Tongshuang Wu, Jordan Boyd-Graber, Daniel S Weld, and Leah Findlater. 2020. No explainability without accountability: An empirical study of explanations and feedback in interactive ml. In *Proceedings of the 2020 CHI Conference on Human Factors in Computing Systems*. 1–13.

[114] Lucy A Suchman. 1987. *Plans and situated actions: The problem of human-machine communication*. Cambridge university press.

[115] The Artificial Intelligence Task Force. 2018. *No Report of Task Force on Artificial Intelligence*. Technical Report. https://dpiit.gov.in/whats-new/report-task-force-artificial-intelligence

[116] Harry Tucker. 2016. Australian Uber drivers say the company is manipulating their ratings to boost its fees. *Business Insider Australia* 20 (2016).

[117] James Vincent. 2016. Twitter taught {Microsoft}'s friendly {AI} chatbot to be a racist asshole in less than a day. https://www.theverge.com/2016/3/24/11297050/tay-microsoft-chatbot-racist

[118] John Vines, Mark Blythe, Stephen Lindsay, Paul Dunphy, Andrew Monk, and Patrick Olivier. 2012. Questionable concepts: critique as resource for designing with eighty somethings. In *Proceedings of the SIGCHI Conference on Human Factors in Computing Systems*. 1169–1178.

[119] Danding Wang, Qian Yang, Ashraf Abdul, and Brian Y Lim. 2019. Designing theory-driven user-centric explainable AI. In *Proceedings of the 2019 CHI conference on human factors in computing systems*. 1–15.

[120] Ruotong Wang, F. Maxwell Harper, and Haiyi Zhu. 2020. Factors Influencing Perceived Fairness in Algorithmic Decision-Making: Algorithm Outcomes, Development Procedures, and Individual Differences. In *Proceedings of the 2020 CHI Conference on Human Factors in Computing Systems* (Honolulu, HI, USA) *(CHI '20)*. Association for Computing Machinery, New York, NY, USA, 1–14. https://doi.org/10.1145/3313831.3376813

[121] Maranke Wieringa. 2020. What to account for when accounting for algorithms: a systematic literature review on algorithmic accountability. In *Proceedings of the 2020 conference on fairness, accountability, and transparency*. 1–18.

[122] Allison Woodruff, Sarah E Fox, Steven Rousso-Schindler, and Jeffrey Warshaw. 2018. A qualitative exploration of perceptions of algorithmic fairness. In *Proceedings of the 2018 chi conference on human factors in computing systems*. 1–14.

[123] Kyra Yee, Uthaipon Tantipongpipat, and Shubhanshu Mishra. 2021. Image Cropping on Twitter: Fairness Metrics, their Limitations, and the Importance of Representation, Design, and Agency. *arXiv preprint arXiv:2105.08667* (2021).